\documentclass[10pt, journal]{IEEEtran}

\usepackage[english]{babel}
\usepackage{tikz, tabularx}
\usetikzlibrary{shapes}
\usepackage[T1]{fontenc}
\usepackage{pifont}
\usepackage{tikz-timing}

\usepackage{array}
\usepackage{epsfig}
\usepackage{epstopdf}
\usepackage{multicol}
\usepackage{multirow}
\usepackage{amsmath}
\usepackage{amssymb}
\usepackage{threeparttable}
\usepackage{bigstrut}
\usepackage{amssymb}
\usepackage[nolist , nohyperlinks]{acronym}
\usepackage{graphicx}
\usepackage{subfigure}
\usepackage{colortbl}
\usepackage{array}
\usepackage{threeparttable}
\usepackage{lipsum}
\usepackage{mathtools}
\usepackage{cite}

\usepackage{algpseudocode}
\usepackage{algorithm}
\usepackage{multicol}
\usepackage[justification=centering]{caption}

\usepackage{lipsum}
\usepackage{blindtext}
\usepackage{scrextend}

\newtheorem{theorem}{Theorem}

\newtheorem{lemma}[theorem]{Lemma}

\DeclareMathOperator{\Tr}{Tr}

\graphicspath{{../figure/}}
\renewcommand{\vec}[1]{\ensuremath{\boldsymbol{#1}}}
\algnewcommand{\LineComment}[1]{\State \(\triangleright\) #1}

\usepackage{color}

\begin{document}

\title{Impact of Relay Cooperation on the Performance of Large-scale Multipair Two-way Relay Networks}
%


\author{\IEEEauthorblockN{Muris Sarajli\'{c},
		Liang Liu,			
		Fredrik Rusek,		
		Farhana Sheikh\IEEEauthorrefmark{2} and
		Ove Edfors} \\
	\IEEEauthorblockA{
		Department of Electrical and Information Technology, Lund University, Sweden \\
		\IEEEauthorrefmark{2}Intel Labs, Hillsboro, OR, USA \\
		Email: muris.sarajlic@eit.lu.se}}
		
		
\maketitle

\begin{abstract}
We consider a multipair two-way relay communication network, where pairs of user devices exchange information via a relay system. 
The communication between users employs time division duplex, with all users transmitting simultaneously to relays in one time slot and relays sending the processed information to all users in the next time slot.
The relay system consists of a large number of single antenna units that can form groups.
Within each group, relays exchange channel state information (CSI), signals received in the uplink and signals intended for downlink transmission.
On the other hand, per-group CSI and uplink/downlink signals (data) are not exchanged between groups, which perform the data processing completely independently.
Assuming that the groups perform zero-forcing in both uplink and downlink, we derive a lower bound for the ergodic sumrate of the described system as a function of the relay group size.
By close observation of this lower bound, it is concluded that the sumrate is essentially independent of group size when the group size is much larger than the number of user pairs.
This indicates that a very large group of cooperating relays can be substituted by a number of smaller groups, without incurring any significant performance reduction.
Moreover, this result implies that relay cooperation is more efficient (in terms of resources spent on cooperation) when several smaller relay groups are used in contrast to a single, large group.
\end{abstract}

\acresetall


\section{Introduction}
\label{sec:introduction}
Multipair two-way relay systems have attracted significant attention in the research community, due to their inherent ability to overcome the halving of system sumrate (stemming from half-duplex operation), with essential doubling of sumrate compared to ordinary one-way relaying \cite{Wang2011}.
Much of the research efforts regarding these systems were focused on developing signal processing algorithms at the relays that are tailored to fit a certain target objective, e.g. interference cancellation or sumrate maximization \cite{Wang2011}, \cite{Zhang2012}.
Recently, multipair two-way relaying systems with a large number of relays/relay antennas were considered \cite{Ngo2013, Jin2015, Le2015, Dai2016, Liu2014, Kong2016}.
By employing a large-scale relay network, system performance is boosted via the channel hardening effect, thus either improving system sumrate or increasing coverage compared to relay systems that operate on a smaller scale.

Previous work on large-scale multipair two-way (LS--MTW) relay systems is limited to two extreme scenarios.
In the first scenario, a large number of non-cooperating single-antenna relays processes the data in a fully decentralized fashion.
Such a setup is described in \cite{Ngo2013}, where individual single-antenna relays perform amplify--and--forward processing of the data on the bidirectional links, based only on their local CSI.
No data or CSI is exchanged between the relays, and their sheer large number is relied upon to deliver satisfying performance.
On the other end of the spectrum is the scenario where a single relay with a large number of antennas performs the uplink and downlink processing.
Theoretical performance characterization for this setup was analyzed in \cite{Jin2015, Le2015, Dai2016, Liu2014, Kong2016}, where the relay is assumed to employ simple linear processing (maximum ratio combining/transmission and zero forcing).

\begin{figure}[h!]
	\centering      
	\includegraphics[width=0.5\textwidth]{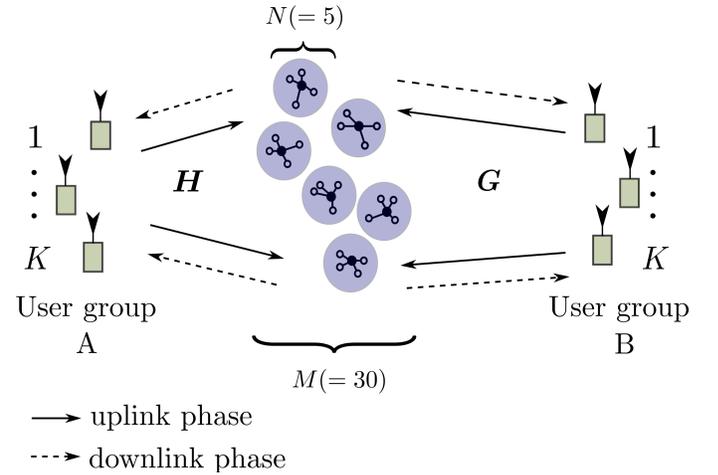}
	\caption{Multipair bidirectional relay network with an arbitrary degree of relay cooperation}
	\label{fig:system_diagram}
\end{figure}

To the best of our knowledge, there is no prior analysis of LS--MTW relay networks with an arbitrary degree of cooperation among relays.
By ``degree of cooperation'' we here refer to the number of relays that will exchange data and channel state information inside a closed group, with no exchange occurring between groups.
The number of closely cooperating relays directly trades system performance against data exchange cost, and is therefore an important design parameter for practical implementations of LS--MTW networks.
An illustration of a MTW network with grouped relays is shown in Fig. \ref{fig:system_diagram}. 

This work provides a comprehensive analysis of the effects of relay cooperation in an LS--MTW relay system where zero-forcing processing is used.
We derive a lower bound for ergodic system sumrate that is tight at high SNR.
Furthermore, we make use of this bound to analyze the behavior of system performance as the number of closely cooperating relays changes.
Finally, we reflect on the choice of the degree of relay cooperation that maximizes the cost-effectiveness of cooperation.


\section{System setup}

In this paper, we analyze the multi-pair bidirectional symmetric relay network, where $M$ relays serve to connect two separate groups of $K$ user units, as illustrated in Fig. \ref{fig:system_diagram}.
Each user from group A is connected in a pairwise fashion with a corresponding user in group B, with $K$ pairs formed in total.
It is assumed that there is no direct link between the users in a pair, so the pairwise connections are established solely via the relays.
Moreover, the information flow between the two users in a pair is assumed symmetric.

The exchange of information is split in two phases, uplink and downlink, observed from the perspective of the relays, which occur in alternating time slots.
In the uplink phase, all $2K$ user units simultaneously transmit to the relays.
The relays process the received signal and send the processed information to the users in the downlink phase, with the direction of information flow swapped compared to the uplink phase (processed information that came on the uplink of channel $\vec{H}$ is transmitted on the downlink of channel $\vec{G}$, and vice versa).

Furthermore, we assume that it is only the relays that have knowledge of the channels $\vec{H}$ and $\vec{G}$.
In practice, the channels can be estimated in the uplink by transmission of orthogonal pilot sequences from the user units, and downlink channels are then automatically obtained assuming that radio channel reciprocity holds and that reciprocity calibration is performed at the relays.
In this sense, the analyzed relay system is equivalent to collocated or distributed massive MIMO (MaMI) systems working in time division duplex (TDD), and channel estimation and reciprocity calibration methods developed for TDD MaMI readily apply \cite{Larsson2014, Vieira2017, Rogalin2014}.
However, for clarity of analysis, in this work we assume perfect channel knowledge and channel reciprocity.

The focus of investigation in this work is the impact of cooperation between the relays on the overall system performance.
The level of inter-relay cooperation is the parameter that trades off network performance with cost of backhaul information exchange.
To this end, we assume the following hierarchical structure of the relay system:
\begin{itemize}
	\item The relays are assumed to be divided in equally-sized groups, each containing $N$ relays.
		  Inside the group, channel state information (CSI), symbols received in the uplink phase and symbols to be transmitted in the downlink phase are shared mutually among all relays.
		  Moreover, the relays inside the group are assumed to be time and frequency synchronized.
		  The data is congregated and processed for uplink and downlink in a central group processor (CGP).
		  One of the relays can take on the role of the CGP, and is referred to as the group master (represented by $\bullet$ in Fig. \ref{fig:system_diagram}).
		  Most importantly, no data or CSI information is exchanged between the groups, and each group performs data processing independently of others.
	\item Groups (or equivalently, group masters) are assumed to be synchronized in time and frequency, and this is the only form of inter-group cooperation.
\end{itemize}

The two-tier hierarchy of cooperation enables us to cover the entire space of cooperative networks that lies in between the two extreme cases:
\begin{itemize}
	\item For $N = 1$ we have the fully decentralized multipair relaying scenario, where single-antenna relays use their local CSI to process and relay the received data without exchanging any CSI or received data information with other relays.
	Such a scenario was analyzed in \cite{Ngo2013}.
	\item The case of $N = M$ represents the perfectly centralized relaying scenario where all CSI and received data is available at a central point that performs data processing.
	Usually this setup is cast in the form of one (massive) MIMO relay, as in \cite{Jin2015}.
\end{itemize}

In general, there are no constraints on the geographical distribution of relays in a group, which can be collocated or distributed.
Likewise, the type of connections between the relays in a group is arbitrary and can be wireless or wired.
We note, however, that a group of $N$ collocated users can be observed as a single MIMO relay with $N$ antennas.
We also note that the stratification of relay cooperation enables the design of a layered and scalable synchronization protocol.
Instead of synchronizing all the relays to a common beacon, synchronization can be done first on the group level and then among the group masters, resulting in a completely decentralized synchronization scheme.
For the sake of simplicity, we assume that the intra- and inter-group synchronization is perfect, and leave the analysis of the impact of synchronization errors as a subject for future work.


\section{System model}

We start the description of the system model by denoting with
\begin{equation}
L = \frac {M}{N}
\end{equation}
the total number of relay groups.
In each channel use, transmitted user symbols are represented by the $2K \times 1$ complex vector $\vec{x}$ with covariance matrix $\mathbb{E} \left \{  \vec{x} \vec{x}^H \right \} = \vec{I}_{2K}^{}$.
The user symbol vector can be represented as $\vec{x} = \left [ \vec{x}_A^T ~ \vec{x}_B^T  \right ]^T$, where $\vec{x}_A^{}$ and $\vec{x}_B^{}$ are symbols transmitted from the left-hand-side and right-hand-side groups of users, illustrated in Figure \ref{fig:system_diagram}, respectively.

Focusing on the $i$th relay group, we build the system model step by step, following the uplink - downlink flow of information.
First, we denote by
\begin{equation}
\vec{\Xi}_{u, i} = \left [ \vec{H}_i  ~ \vec{G}_i  \right ]_{N \times 2K}
\end{equation}
the composite uplink channel between the $i$th relay group and all the users.
Received signal vector at the $i$th relay group is then
\begin{equation}
\vec{y}_i = \sqrt {P_U^{}} \vec{\Xi}_{u, i} \vec{x} + \vec{n}_{R, i}^{},
\end{equation}
where $\vec{n}_{R, i}^{}$ is an $N \times 1$ zero-mean circularly symmetric complex Gaussian (ZMCSCG) vector of thermal noise with covariance matrix $\mathbb{E} \left \{ \vec{n}_{R, i}^{} \vec{n}_{R, i}^H  \right \} = N_{0, R}^{} \vec{I}_N^{}$ and $P_U^{}$ is the uplink transmit power per user, assumed to be same for all users.

The received signal in the uplink is linearly filtered with $\vec{W}_{u, i}$ to yield the estimates of user data symbols:
\begin{equation}
\label{eq:user_data_estimates}
\hat{\vec{x}}_i = \vec{W}_{u, i} \vec{y}_i = \sqrt {P_U^{}} \vec{W}_{u, i} \vec{\Xi}_{u, i} \vec{x} + \vec{W}_{u, i} \vec{n}_{R, i}^{}.
\end{equation}

After uplink filtering, the downlink precoder $\vec{W}_{d, i}$ is applied to symbol estimates, together with a scaling factor $\alpha_i$ ensuring proper transmitted power.
We assume that the user power allocation in the downlink is uniform.

The uplink/downlink linear processing can be compactly represented by a general complex gain matrix
\begin{equation}
\label{eq:composite_processing_matrix}
\vec{W}_i = \vec{W}_{d, i} \vec{W}_{u, i}.
\end{equation}
Altogether, the transmit signal vector from the $i$th relay group is
\begin{align}
\label{eq:transmitted_signal_relays}
\vec{t}_i &= \alpha_i \vec{W}_{d, i} \hat{\vec{x}}_i \\ \nonumber
&= \alpha_i \sqrt {P_U^{}} \vec{W}_i \vec{\Xi}_{u, i} \vec{x} + \alpha_i \vec{W}_i \vec{n}_{R, i}^{}.
\end{align}

The power scaling coefficient $\alpha_i$ is determined so that the transmitted power per group averaged over data and noise realizations equals $P_{R, i}^{}$.
For the heavily restricted decentralized setup considered here, a practically implementable strategy of power allocation between relay groups is that all groups have the same transmit power, so $P_{R, i}^{} = P_{R}^{}, \forall i$.
Overall, we have
\begin{equation}
\mathbb{E}_{\vec{x}, \vec{n}}  \left \{  \left | \left |   \vec{t}_i     \right |   \right |^2 \right \}  = P_R^{},
\end{equation}
which readily yields
\begin{equation}
\label{eq:alpha_def}
\alpha_i = \sqrt {  \frac {P_R}{P_U^{}  \left |  \left | \vec{W}_i \vec{\Xi}_{u, i}  \right |  \right |_F^2 + N_{0, R}^{}  \left |  \left | \vec{W}_i \right |  \right |_F^2 }}.
\end{equation}

We define the composite downlink channel between the $i$th relay group and all the users as
\begin{equation}
\vec{\Xi}_{d, i}^T = \left [ \vec{G}_i  ~ \vec{H}_i  \right ]_{N \times 2K}^T
\end{equation}

The contribution of the $i$th relay group to the received signal at the users, $\vec{z}_i$, is thus
\begin{align}
\vec{z}_i &= \begin{bmatrix}
z_{B, i}^{} \\
z_{A, i}^{}
\end{bmatrix} =	
\vec{\Xi}_{d, i}^T \vec{t}_i^{} \\ \nonumber
&= \alpha_i^{} \sqrt {P_U^{}} \vec{\Xi}_{d, i}^T \vec{W}_i^{} \vec{\Xi}_{u, i}^{} \vec{x} + \alpha_i^{} \vec{\Xi}_{d, i}^T \vec{W}_i^{} \vec{n}_{R, i}^{}.
\end{align}

The total received signal vector at the users is hence
\begin{align}
\label{eq:complete_system_model}
&\vec{z} = \sum_{i=1}^L \vec{z}_i + \vec{n}_U^{} \\ \nonumber
&= \sqrt {P_U^{}} \left (  \sum_{i=1}^L \alpha_i^{} \vec{\Xi}_{d, i}^T \vec{W}_i^{} \vec{\Xi}_{u, i}^{} \right ) \vec{x} \\ \nonumber 
&+ \sum_{i=1}^L \alpha_i^{} \vec{\Xi}_{d, i}^T \vec{W}_i^{} \vec{n}_{R, i}^{} + \vec{n}_U^{},
\end{align}
where $\vec{n}_U^{}$ is the $N \times 1$ ZMCSCG vector of thermal noise at the users, with covariance $\mathbb{E} \left \{ \vec{n}_U^{} \vec{n}_U^H  \right \} = N_{0, U}^{} \vec{I}_{2K}^{}$.

For the benefit of further analysis, we define the uplink and downlink SNRs as
\begin{equation}
\mathit{SNR}_u = \frac {P_U^{}}{N_{0, R}^{}},~\text{and}~\mathit{SNR}_d = \frac {P_R^{}}{N_{0, U}^{}}.
\end{equation}

The overall system model \eqref{eq:complete_system_model} can be expanded for the received symbol at a particular user, say $k$th user from group A.
This reveals that the performance in the general case is limited by four distinct impairments: self-interference, interuser interference, precoded thermal noise at the relays and thermal noise at the users:
\begin{align}
\label{eq:broken_down_system_model}
&z_{A, k}^{} = \underbrace{ \sqrt{P_U^{}} \left (  \sum_{i=1}^L \alpha_i^{} \vec{h}_{k, i}^T \vec{W}_i^{} \vec{g}_{k, i}^{} \right ) x_{B, k}^{}}_{\text{wanted information},~x_{W, k}^{}} \\ \nonumber
&+ \underbrace{\sqrt{P_U^{}} \left (  \sum_{i=1}^L \alpha_i^{} \vec{h}_{k, i}^T \vec{W}_i^{} \vec{h}_{k, i}^{} \right ) x_{A, k}^{}}_{\text{self-interference},~\nu_{\textit{SI}, k}^{}} \\ \nonumber
&+ \underbrace{\sqrt{P_U^{}} \sum_{i=1}^L \alpha_i^{} \sum_{\substack{j = 1, \\ j \neq k}}^K  \vec{h}_{k, i}^T \vec{W}_i^{} \left ( \vec{h}_{j, i}^{} x_{A, j}^{} + \vec{g}_{j, i}^{} x_{B, j}^{} \right )}_{\text{interuser interference},~\nu_{\textit{IUI}, k}^{}} \\ \nonumber
&+ \underbrace{\sum_{i=1}^L \alpha_i^{} \vec{h}_{k, i}^T \vec{W}_i^{} \vec{n}_{R, i}^{}}_{\text{precoded noise from relays},~\nu_{\textit{PN}, k}^{}} + \underbrace{n_{A, k}^{}}_{\text{thermal noise at users}}.
\end{align}  
In the follow-up, we consider the case when zero-forcing is chosen as the linear processing scheme at individual relay groups and analyze system performance, averaged over channel realizations.
The goal of the analysis is to determine the closed-form dependence of system performance (quantified by ergodic system sumrate) on relay group size $N$.

\section{Ergodic system sumrate calculation with per-group zero-forcing}

If zero-forcing (ZF) is chosen for linear processing, the uplink and downlink processing matrices at each relay group are calculated as
\begin{align}
\label{eq:ZF_processing_def}
\vec{W}_{u, i} &= \left ( \vec{\Xi}_{u, i}^H \vec{\Xi}_{u, i}^{}  \right )^{-1} \vec{\Xi}_{u, i}^H,~\text{and} \\ \nonumber
\vec{W}_{d, i} &= \vec{\Xi}_{d, i}^* \left ( \vec{\Xi}_{d, i}^T \vec{\Xi}_{d, i}^*  \right )^{-1}.
\end{align}
Back-to-back ZF processing will completely eliminate self- and interuser interference, leaving the precoded noise from relays and noise at the user terminals as sources of impairment.
A strong requirement for total interference elimination is that $N > 2K$.

In order to gain some insight in the connection between system performance (quantified by system sumrate) and system parameters $M$ and $N$, we assume that $\mathit{SNR}_u$ is high.
Typically, a high SNR would mean that the geographical distances between users and relays are small, and that the relays are used to boost the system sumrate (in contrast to e.g. a range extension scenario).

Under the high-SNR assumption and additionally assuming that $\vec{H}_i$ and $\vec{G}_i$ are well-conditioned, the influence of precoded thermal noise at the relays can be neglected, so the system model \eqref{eq:complete_system_model} simplifies to
\begin{equation}
\label{eq:high_SNR_system_model}
\vec{z} = \sqrt {P_U^{}} \sum_{i=1}^L \alpha_i \vec{x} + \vec{n}_U^{}.
\end{equation}

A basis for performance evaluation is the per-user $\mathit{SNR}$, defined for the $k$th user in group A as the ratio of powers of the information signal part and impairments from \eqref{eq:broken_down_system_model}, which, due to all interference being eliminated and the high-SNR assumption, becomes
\begin{align}
	\mathit{SNR}_{A, k}^{} &= \frac {\mathbb{E}_x \left \{ \left | x_{W, k} \right |^2  \right \}}{N_{0, U}^{}} 
						    = \frac {P_U^{}}{N_{0, U}^{}} \left ( \sum_{i=1}^L \alpha_i \right )^2 \\ \nonumber
						   &\stackrel{\mathit{a)}}{=} \frac {P_R^{}}{N_{0, U}^{}} \left ( \sum_{i=1}^L \frac {1}{\sqrt{\left |  \left | \vec{W}_i \vec{\Xi}_{u, i}  \right |  \right |_F^2}}  \right )^2 \\ \nonumber
						   &\stackrel{\mathit{b)}}{=} \frac {P_R^{}}{N_{0, U}^{}} \left ( \sum_{i=1}^L \frac {1}{\sqrt{\Tr \left [ \left ( \vec{\Xi}_{d, i}^T \vec{\Xi}_{d, i}^*  \right )^{-1}  \right ]}}  \right )^2,
\end{align}
where $\mathit{a)}$ follows from \eqref{eq:alpha_def}, with the assumption of high-SNR at the relays, and $\mathit{b)}$ from \eqref{eq:composite_processing_matrix} and \eqref{eq:ZF_processing_def}.

Instantaneous per-user performance is characterized by user rate:
\begin{equation}
	R_{A, k} = \log_2 \left ( 1 + \mathit{SNR}_{A, k}^{} \right ) \quad \left [ \text{bps/Hz} \right ],
\end{equation}
and overall system performance by ergodic system sumrate, calculated as
\begin{equation}
\label{eq:sumrate_def}
R = \mathbb{E}_{\vec{H}, \vec{G}} \left \{ \frac {1}{2} \left ( 2 \sum_{k=1}^K  R_{A, k}  \right ) \right \} =  \sum_{k=1}^K \mathbb{E}_{\vec{H}, \vec{G}} \left \{ R_{A, k}  \right \}.
\end{equation}
The factor of 2 accounts for the fact that the information flow in the system is symmetric, so members of the $k$th user pair have the same information transmission capacity.
The factor of $1/2$, on the other hand, stems from half-duplex operation.
From here, the benefit of the symmetric multi-pair setup compared with ordinary relaying schemes becomes clear: simultaneous and symmetric transmission from both user groups manages to (approximately) compensate for the halving of the capacity due to TDD splitting of uplink and downlink.

We proceed with calculating \eqref{eq:sumrate_def}, and in the process, we make use of
\begin{lemma}
	\label{lemma_Jensen}
	Let $\vec{\Psi} = \left [ \Psi_1~\Psi_2~\dots~\Psi_N \right ]$ be a vector of nonnegative random variables $\Psi_i$, with $\psi_i$ denoting realizations of $\Psi_i$.
	Then
	\begin{equation*}
		\mathbb{E}_{\vec{\Psi}} \log_2 \left [ 1 + \left ( \sum_{i} \frac {1}{\sqrt{\psi_i}}  \right )^2  \right ]  > \log_2 \left [ \left ( \sum_{i} \frac {1}{\sqrt{\mathbb{E}_{\vec{\Psi}}  \psi_i  }}  \right )^2 \right ].
	\end{equation*}
\end{lemma}

\textit{Proof}: The proof is given in Appendix A.

We employ Lemma 1, assuming in the process that $P_R^{}/N_{0, U}^{} = 1$ without loss of generality, to obtain a lower bound on the ergodic information rate of the $k$th user as
\begin{equation}
\label{eq:ZF_user_rate_1}
	\mathbb{E}_{\vec{H}, \vec{G}} \left \{ R_{A, k}^{} \right \} > \log_2 \left [ \frac {P_R^{}}{N_{0, U}^{}} \left ( \sum_{i=1}^L \frac {1}{\sqrt{\mathbb{E}_{\vec{H}, \vec{G}} \left \{ \zeta_i \right \}}}  \right )^2 \right ],
\end{equation}
where, for sake of clarity, we introduce the ZF precoding scaling factor
\begin{equation}
\zeta_i = \Tr \left [ \left ( \vec{\Xi}_{d, i}^T \vec{\Xi}_{d, i}^*  \right )^{-1} \right ].
\end{equation}

Now we assume that $\vec{H}_i$ and $\vec{G}_i$ are iid Rayleigh fading channels with pathloss and shadowing, modeled as
\begin{equation}
\label{eq:channel_model_fundamental}
 \vec{H}_i = \widetilde{\vec{H}}_i \vec{D}_{A, i}^{1/2}~\text{and}~ \vec{G}_i = \widetilde{\vec{G}}_i \vec{D}_{B, i}^{1/2}.
\end{equation} 
The entries of $N \times K$ matrices $\widetilde{\vec{H}}$ and $\widetilde{\vec{G}}$ are iid ZMCSCG with unit variance, and the diagonal matrices
\begin{align}
	\vec{D}_{A, i}^{1/2} &= \text{diag} \left ( \sqrt{\beta_{A, 1, i}}, \sqrt{\beta_{A, 2, i}}, ~\dots~\sqrt{\beta_{A, K, i}} \right )~\text{and} \\ \nonumber
	\vec{D}_{B, i}^{1/2} &= \text{diag} \left ( \sqrt{\beta_{B, 1, i}}, \sqrt{\beta_{B, 2, i}}, ~\dots~\sqrt{\beta_{B, K, i}} \right )									   
\end{align}
are used to model propagation losses and large-scale fading.
In order for channel matrices to be decomposable as in \eqref{eq:channel_model_fundamental}, the propagation amplitude gain $\sqrt{\beta_{(A, B), k, i}} > 0$ needs to be the same from user $k$ to all relays in group $i$, which implies that the relays of that group are assumed to be collocated and experience the same large scale fading in relation to user $k$.
These conditions are readily satisfied if a relay group is implemented in form of a single MIMO relay with a compact form factor.
Otherwise, they can be met by applying an appropriate relay grouping scheme, which is an interesting research problem in itself, but falls outside of the scope of this paper.

Using well-known results from random matrix theory \cite{Tulino2004} and the identity $\Tr (\vec{A} \vec{B} ) = \Tr ( \vec{B} \vec{A})$, it can be shown that
\begin{equation}
\label{eq:expectation_zeta}
	\mathbb{E}_{\vec{H}, \vec{G}} \left \{ \zeta_i \right \} = \frac {\gamma_i}{N - 2K},
\end{equation}
where
\begin{equation}
\label{eq:definition_gamma}
	\gamma_i = \sum_{k=1}^K \left ( \frac {1}{\beta_{A, k, i}} + \frac {1}{\beta_{B, k, i}} \right ).
\end{equation}
Combining \eqref{eq:expectation_zeta} with \eqref{eq:ZF_user_rate_1} yields the lower bound on per-user rate
\begin{equation}
\label{eq:user_rate_lower_bound}
	\mathbb{E}_{\vec{H}, \vec{G}} \left \{ R_{A, k}^{} \right \} > \log_2 \left [ \frac {P_R^{}}{N_{0, U}^{}} (N - 2K) \delta \right ],
\end{equation}
with
\begin{equation}
\label{eq:definition_delta}
	\delta = \left ( \sum_{i=1}^L \frac {1}{\sqrt{\gamma_i}} \right )^2.
\end{equation}
For convenience of exposition, in the follow-up we will refer to $\gamma_i$ and $\delta$ as power imbalance factor and array gain degradation factor, respectively.

Overall, the lower bound on system sumrate for the multipair two-way relay system with relay grouping and ZF processing at high SNR is given from \eqref{eq:sumrate_def} and \eqref{eq:user_rate_lower_bound} by
\begin{equation}
\label{eq:sumrate_final_general}
 	R > \max \left \{0, K \log_2 \left [ \frac {P_R^{}}{N_{0, U}^{}} \left ( N - 2K  \right ) \delta \right ] \right \}.
\end{equation}

\section{Analysis and discussion}

In order to have a fair comparison between systems, we assume that the pathloss and shadowing power gains are normalized so that
\begin{equation}
\label{eq:normalized_pathloss_constraint}
\Tr \left ( {\vec{D}_{A, i}} \right ) = \Tr \left ( {\vec{D}_{B, i}} \right ) = K,~\forall i,
\end{equation}
which implies $\mathbb{E} \left \{  \left |  \left |  \vec{H}_i  \right |  \right |_F^2 \right \} = \mathbb{E} \left \{  \left |  \left |  \vec{G}_i  \right |  \right |_F^2 \right \} = N K, \forall i$.
Given the constraint \eqref{eq:normalized_pathloss_constraint}, it is easy to show that power imbalance and array gain degradation factors are lower (respectively, upper) bounded as
\begin{equation}
\label{eq:gamma_delta_bounds}
\gamma_i \geq 2K~\text{and}~\delta \leq \frac {L^2}{2K},
\end{equation}
where equality holds in the case ${\vec{D}_{A, i} = \vec{D}_{B, i} = \vec{I}_K}$.
In other words, the lower bound on system sumrate from \eqref{eq:sumrate_final_general} is maximized when there are no pathloss/shadowing power imbalances between users.
In practical system deployments, such imbalances will invariably exist and the resulting degradation of sumrate can be combatted by either performing waterfilling-based user power weighting in the downlink, or by employing advanced user scheduling techniques.
Analysis of the effects of these approaches is beyond the scope of this work.

For a fair comparison between relay systems with differing $M$ and $N$, we need to assume that the amount of transmit power allocated to the entire relay system is fixed, and we denote this power by $P_T^{}$.
As mentioned previously, due to limited coordination, it is reasonable to assume that the total power allocated to relays is distributed equally among relay groups, so $P_R^{} = P_T^{} / L$.
By taking into account \eqref{eq:gamma_delta_bounds}, we can write ${\delta = \frac {L^2}{2K} \epsilon,~\epsilon \leq 1}$, which yields the lower bound on sumrate that allows for a fair comparison between different relay systems:
\begin{equation}
\label{eq:sumrate_final_fair}
R > \max \left \{0, K \log_2 \left ( \frac {P_T^{}}{N_{0, U}^{}} \frac {M}{N} \frac {N - 2K}{2K} \epsilon \right ) \right \}.
\end{equation}

Now we can consider the case when $N \gg 2K$.
Even for a large number of user pairs, this case is feasible due to the fundamental assumption of a large number of relay units, $M \gg 1$.
The array gain term from \eqref{eq:sumrate_final_fair} then becomes
\begin{equation}
	\frac {M}{N} \frac {N - 2K}{2K} \epsilon \approx \frac {M}{2K} \epsilon,
\end{equation}
that is, the array gain becomes independent of $N$.
This insight is of fundamental importance for practical deployments of LS-MTW systems.
What it implies is that, in the regime with a large number of relays $M$, tight cooperation in information processing between all $M$ relays is not necessary.
Instead, small, independent groups of tightly cooperating relays can be formed, and such a setup experiences only a marginal degradation of system sumrate compared to the case when all relays are cooperating.
If we substitute the notion of a tightly cooperating relay group with a more specific notion of a MIMO relay, we can conclude that a single massive MIMO relay performing ZF can be substituted with several simpler and cheaper MIMO relays with smaller numbers of antennas, with a negligible reduction in system performance.
\begin{figure}[h!]
	\centering      
	\includegraphics[width=0.5\textwidth]{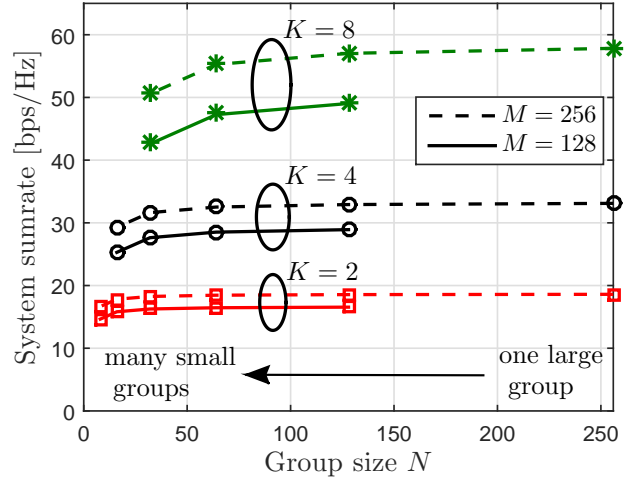}
	\caption{Theoretical and simulated ergodic sumrate of a LS-MTW system with ZF in uplink and downlink.  Markers: simulation results, full and dashed lines: theoretical lower bounds. Minimum theoretical relay group size = $2K + 1$, maximum = $M$. $\mathit{SNR}_u = \mathit{SNR}_d = 10$ dB and no power imbalance assumed. }
	\label{fig:sumrate_comparison_no_pathloss}
\end{figure}

The presented observations are corroborated by simulations, results of which are presented in Fig. \ref{fig:sumrate_comparison_no_pathloss}, where the lower bound \eqref{eq:sumrate_final_fair} is compared to simulated system sumrate, averaged over channel realizations.
It is assumed that the total transmit power in the system, which we denote by $P_{\mathit{tot}}$, is split between users and relays in two equal parts, so $P_U = P_{\mathit{tot}}/4K$ and ${P_R = P_{\mathit{tot}}/2L}$.
The results show an excellent match between the theoretical lower bound \eqref{eq:sumrate_final_fair} and simulations.
Moreover, it is clearly demonstrated how substituting one large group ($N = M$) with several smaller and independent groups of relays introduces only a slight degradation of sumrate (in the most extreme cases, sumrate degrades by $10.5\%$ to $12.5\%$ for the setups considered).

In order to gain deeper understanding of tradeoffs encountered in the design of LS-MTW systems, in addition to sumrate, we also need to take into account the cost of enabling cooperation between the relays.
This cost, which we denote by $C$, quantifies the resources spent (e.g. energy, bandwidth) or penalties in system performance incurred (e.g. latency) when CSI and uplink/downlink data are exchanged inside a relay group.
In particular, we focus on resources that can be \emph{reused} between groups.
An example system setup would feature relays inside a group exchanging CSI and data with the CGP over dedicated short-range wireless links and employing frequency division multiplexing (FDM).
With enough physical separation between individual groups, the short range of intra-group backhaul links would mean that the bandwidth dedicated for cooperation can be reused between groups.
Moreover, the use of FDM implies that this bandwidth is proportional to the group size:
\begin{equation}
C = c_{\textnormal{BW}}^{} N~[\textnormal{Hz}],
\end{equation}
where $c_{\textnormal{BW}}^{}$ is the bandwidth of the frequency slot allocated for one user-CGP link.
As discussed previously, for ${N \gg 2K}$, sumrate is independent of $N$.
Therefore, the cooperation efficiency of the described system,
\begin{equation}
\eta =  \frac{R}{C},
\end{equation}
increases for decreasing $N$ when $N$ is large.
\begin{figure}[h!]
	\centering      
	\includegraphics[width=0.5\textwidth]{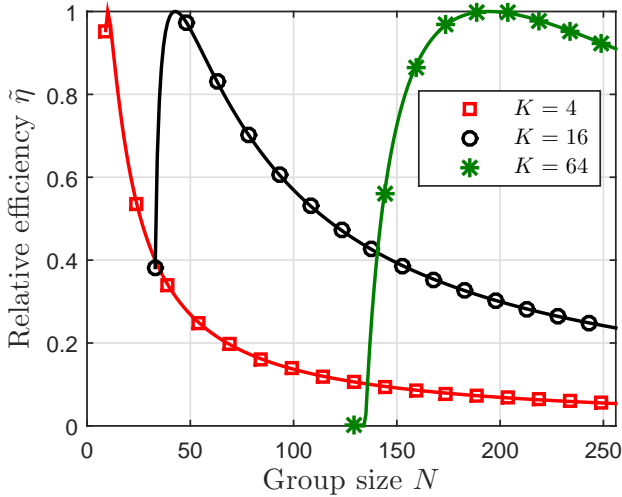}
	\caption{Relative cooperation efficiency with reusable cooperation resources. $\mathit{SNR}_u = \mathit{SNR}_d = 10$ dB, $M = 256$.}
	\label{fig:efficiency}
\end{figure}

Relative cooperation efficiency function $\tilde{\eta} = \eta/\max{ \left \{  \eta \right \} }$ is shown in Fig. \ref{fig:efficiency}, where $N \in \mathbb{N}$ but the constraint $L \in \mathbb{N}$ is relaxed.
These results support the notion that using one large cooperating group is a suboptimal strategy from the point of view of cooperation efficiency, especially for low values of K.
\section{Conclusion}
We have analyzed a multipair two-way relay system with a large number of relays.
The relays are assumed to form groups inside which data and channel state information is exchanged, and processing is done independently from other groups.
Assuming that the groups perform zero-forcing and that the SNR is large, we derive a closed-form expression for a tight lower bound on the system sumrate.
An asymptotic analysis of the bound shows that the sumrate is essentially independent from group size $N$ when $N \gg 2K$.
This implies that one large group of cooperating users can be substituted with several smaller groups, with no significant impact on performance.
We extend this result to take into account the efficiency of information exchange that supports relay cooperation.
It is shown that using several smaller relay groups is more efficient than the use of a single large group, if the resource used for intra-group information exchange is reusable between groups.

\section*{Acknowledgment} 
\addcontentsline{toc}{section}{Acknowledgment} 
The authors would like to thank the Intel Corporation for providing the funds for the research, which was conducted as a part of the SRC MSR-Intel Research Project P28832, ``Coordination in Distributed Multi-User High-Performance Dense Networks''.

\newpage
\section*{Appendix A: Proof of Lemma 1}
\label{app:Appendix_A}
By taking into account Jensen's inequality
\begin{equation*}
\mathbb{E} f(X) \geq (\leq)~f\left ( \mathbb{E} X \right ),
\end{equation*}
$f(X)$ convex (concave), we can form a chain of (in)equalities
\begin{align*}
	&\mathbb{E} \log_2 \left [ 1 + \left ( \sum_{i} \frac {1}{\sqrt{\psi_i}}  \right )^2  \right ]  > \frac {1}{\ln {2}} \mathbb{E}  \ln \left [ \left ( \sum_{i} \frac {1}{\sqrt{\psi_i}}  \right )^2  \right ] \\ \nonumber
	&= \frac {2}{\ln{2}} \mathbb{E}  \ln \left ( \sum_i e^{-\frac {1}{2} \ln \psi_i} \right ) \stackrel{\textit{a)}}{\geq} \frac {2}{\ln 2} \ln \left ( \sum_i e^{-\frac{1}{2} \mathbb{E}  \ln  \psi_i  }  \right ) \\ \nonumber
	&\stackrel{\textit{b)}}{\geq} \frac {2}{\ln 2} \ln \left ( \sum_i e^{-\frac{1}{2} \ln \mathbb{E}  \psi_i  }  \right ) = \log_2 \left [ \left ( \sum_{i} \frac {1}{\sqrt{\mathbb{E}  \psi_i  }}  \right )^2 \right ],
\end{align*}
where inequality \textit{a)} follows from convexity of $\ln \sum_i e^{y_i}$ on $\mathbb{R}^n$, so 
\begin{equation*}
	\mathbb{E} \ln \sum_i e^{y_i}\geq \ln \sum_i e^{\mathbb{E}  y_i  }.
\end{equation*}
Inequality \textit{b)} follows from concavity of $\ln()$ and from the fact that $e^{-z}$ is monotonically decreasing, which yields
\begin{equation*}
	e^{-\frac {1}{2} \mathbb{E}  \ln \psi_i  } \geq e^{-\frac {1}{2}  \ln \mathbb{E} \psi_i  }.
\end{equation*}

%
%
%

\bibliographystyle{IEEEtran}
\bibliography{dense_nw_overview}{}  
\input{dense_nw_overview.acro}

\end{document}